\documentstyle[12pt,aaspp4]{article}
\makeatletter
\def\lnyoro{\mathrel{\mathpalette\gl@align<}}
\def\gnyoro{\mathrel{\mathpalette\gl@align>}}
\def\gl@align#1#2{\lower.6ex\vbox{\baselineskip\z@skip\lineskip\z@\ialign{$\m@th
#1\hfil##\hfil$\crcr#2\crcr\sim\crcr}}}
\makeatother

\begin{document}

\title{\bf GLOBAL LAW FOR DUST-TO-GAS RATIO OF SPIRAL GALAXIES}
\author{\bf HIROYUKI HIRASHITA$^{1,2}$}
\affil
{$^1$:  Department of Astronomy, Faculty of Science, Kyoto University,
Sakyo-ku, Kyoto 606-8502, Japan}
\affil
{$^2$:  Research Fellow of the Japan Society for the Promotion of
Science}
\centerline{Nov. 9th, 1998 (revised version)}
\centerline{email: hirasita@kusastro.kyoto-u.ac.jp}
\authoremail{hirasita@kusastro.kyoto-u.ac.jp}
\begin{abstract}

Using a one-zone model for the evolution of dust
in spiral galaxies and
applying the instantaneous recycling approximation
to the model equations, we investigate dust-to-gas ratio of spiral
galaxies. Four processes are considered; dust formation from heavy
elements ejected by stellar mass loss, dust destruction in supernova
remnants, dust destruction in star-forming regions, and accretion of
heavy elements onto preexisting dust grains. The equations describe
a simplified relation between dust-to-gas ratio and metallicity.
The relation is independent of star formation rate.
By comparing the theoretical relation with the observational 
data of nearby spiral galaxies, we show (i) that the accretion
process onto
the preexisting dust particles must be taken into account for the
spiral galaxies; (ii) that the efficiency of dust production from 
heavy elements ejected by stars can be constrained by the spiral
galaxies with low metallicity; (iii) that the Salpeter and Scalo
initial mass functions are both consistent with the data of the
spiral galaxies.

\end{abstract}


\section{INTRODUCTION}

Interstellar dust is composed of heavy elements made and ejected
by stars. Supernovae (SNe) govern the
dust content in galaxies: Observationally, dust has been
clearly detected in the SN1987a event (Moseley et al. 1989;
see also Kozasa, Hasegawa, \& Nomoto 1989), and theoretically,
Dwek \& Scalo (1980) demonstrated that SNe are the dominant
sources for the formation of dust grains.
%Thus, dust is formed in star-forming galaxies.
Dwek \& Scalo (1980)
also showed that the dust is destroyed in SN remnants (SNRs).

Recently, galaxy-evolution models including the
evolution of dust content have been developed (Wang 1991;
Takagi \& Arimoto 1997;
Lisenfeld \& Ferrara 1998, hereafter LF98; Dwek
1998, hereafter D98). The processes of dust formation and
destruction by
SNe are taken into account in LF98, in order to
explain the relation between dust-to-gas mass ratio and
metallicity of dwarf irregular galaxies and blue
compact dwarf galaxies. We can find detailed mathematical frameworks
to calculate the dust abundance in any galactic system in D98,
in which the accretion of heavy elements onto
preexisting dust grains is considered in addition to the
processes considered in LF98.

In this paper, we examine the dust-to-gas ratio of nearby
star-forming galaxies with a simple model. We use spiral galaxies
as the sample of star-forming galaxies. %This study will provide
%a basis for understanding of dust in the high-redshift Universe.
First of all, in the next section,
we apply instantaneous recycling approximation to the model
equations in D98 to describe a simplified relation between
dust-to-gas ratio and metallicity. Our model is a one-zone
model (i.e., the spatial distribution of gas, metal, and dust within
a galaxy is not taken into account).
The result is compared with observational data in the
same section.
Finally, we present  summary and conclusion in \S 3.

\section{MODEL}

%In order to investigate the dust content in spiral galaxies, we
%here describe the dust formation and destruction
%processes based on the models in LF98 and D98.

\subsection{Basic Equations}

We, here, present the equations which describe the  changing
rate of the mass of gas, metal and dust in a galaxy.
Here, we adopt
a simple one-zone model, because we are interested in
the {\it global} properties of galaxies. For the model treating the
radial distribution of gas, element, and dust in a galaxy, see D98.
After some modifications, equations (6)--(8)
of LF98 are written as
\begin{eqnarray}
\frac{dM_{\rm g}}{dt} & = & -\psi +E,\label{basic1} \\
\frac{dM_i}{dt}       & = & -X_i\psi +E_i, \\
\frac{dM_{{\rm d}i}}{dt} & = & f_{{\rm in}i}E_i-\alpha f_iX_i\psi
+\frac{M_{{\rm d}i}(1-f_i)}{\tau_{\rm acc}}-
\frac{M_{{\rm d}i}}{\tau_{\rm SN}}.
\label{basic3}
\end{eqnarray}
Here, $M_{\rm g}$ is the mass of gas. The metal is labeled by
$i$ ($i={\rm O}$, C, Si, Mg, Fe, $\cdots$), and $M_{i}$ and
$M_{{\rm d}i}$ denote the total mass of the metal $i$ (in
gas and dust phases) and the mass of the metal $i$ in the dust
phase, respectively.  The star formation
rate is denoted by $\psi$; $E$ is the total injection rate of
mass from stars; $X_i$ is the mass fraction of the element $i$
(i.e., $X_i\equiv M_i/M_{\rm g}$); $E_i$ is the total
injection rate of element $i$ from stars; $f_i$ is the mass fraction
of the element $i$ locked up in dust (i.e., $f_i=M_{{\rm d}i}/M_i$). 
The meanings of the other parameters in the above equations are as
follows: $f_{{\rm in}i}$ represents the dust mass fraction in the
injected material, in other words, the dust condensation efficiency
in the ejecta\footnote{In this formalism, we assume that this
efficiency in stellar winds is the same as that in SN
ejecta. The different efficiencies in the various ejecta are treated
in D98.}; $\alpha$ refers to the efficiency of dust
destruction during star formation ($\alpha =1$ corresponds to
destruction of
only the dust incorporated into the star, and $\alpha >1\, [\alpha
<1]$ corresponds to a net destruction [formation] in the star
formation); $\tau_{\rm acc}$ is the accretion timescale of the
element $i$ onto preexisting dust grains in molecular clouds;
$\tau_{\rm SN}$ is the timescale of dust destruction
by SN shocks. We hereafter assume that only dust
incorporated into stars is destroyed when the stars form; i.e.,
$\alpha =1$.

Our equations above are different from the equations in LF98 in
the following two ways.
The one is that we neglect the outflow term (i.e., we set $W=0$ in
the equations of LF98), since
we expect that outflows of gas from spiral galaxies, in which we
are interested, are not
effective because
their gravitational potentials are much deeper than those of
dwarf galaxies, which LF98 treated. For this reason
and for simplicity, we here neglect the outflow terms.
The other is the addition of the third term of the
left-hand side in equation (\ref{basic3}). This term represents the
accretion of element $i$ onto the preexisting dust grains.
The process is effective in molecular clouds, where the
density of gas is high enough for the dust particles to grow through
collisions with the atoms of heavy elements. The timescale
$\tau_{\rm acc}$ is determined from the lifetimes of molecular
clouds. The detailed discussion about the accretion
process can be found in \S 7 of D98.

Here, we adopt
the instantaneous recycling approximation  (Tinsley
1980): Stars less massive than $m_{\rm l}$
live forever and the others die instantaneously.
This approximation allows us to write $E$ and $E_i$, respectively, as
$E={\cal R}\psi$ and
$E_i=({\cal R}X_i+{\cal Y}_i)\psi$,
where %$X_i$ is the mass fraction of the given heavy element $i$,
${\cal R}$ is the returned fraction of the mass that has formed
stars which is subsequently ejected into the interstellar space,
and ${\cal Y}_i$ is the mass fraction of the element $i$ newly
produced and ejected by stars.\footnote{${\cal R}=R$ and
${\cal Y}_i=y(1-R)$ for the notations in LF98.}
${\cal R}$ and ${\cal Y}_i$ can be obtained using the following
formulae (Maeder 1992):
\begin{eqnarray}
{\cal R} & = & {\int_{m_{\rm l}}^{m_{\rm u}}(m-w_m)\phi(m)dm},
%{\int_{m_{\rm l}}^{m_{\rm u}}m\phi(m)dm},
\label{eqr} \\
{\cal Y}_i & = &
{\int_{m_{\rm l}}^{m_{\rm u}}mp_i(m)\phi(m)dm},
%{\int_{m_{\rm l}}^{m_{\rm u}}m\phi(m)dm},
\label{eqy}
\end{eqnarray}
In equation (\ref{eqr}), $\phi (m)$ is the initial mass function
(IMF), and $m_{\rm u}$ is the upper mass cutoff of stellar mass.
The IMF is normalized so that integral of $m\phi (m)$ in the full
range of stellar mass becomes 1. Thus, $\phi (m)$ has
dimension of the inverse square of mass. In equation (\ref{eqy}),
$w_m$ is the remnant mass
($w_m=0.7M_\odot$ for $m<4M_\odot$ and $w_m=1.4M_\odot$ for
$m>4M_\odot$) and  $p_i(m)$ is the mass fraction of a star of mass
$m$
converted into the element $i$ and ejected. %in a star of mass $m$.
%We use a power-law form of the IMF: $\phi(m)\propto m^{-x}$.

Using the above parameters ${\cal R}$ and ${\cal Y}_i$,
equations (\ref{basic1})--(\ref{basic3}) become
\begin{eqnarray}
\frac{1}{\psi}\frac{dM_{\rm g}}{dt} & = & -1+{\cal R},
\label{basic4} \\
\frac{M_{\rm g}}{\psi}\frac{dX_i}{dt} & = & {\cal Y}_i,
\label{basic5} \\
\frac{M_{\rm g}}{\psi}\frac{d{\cal D}_i}{dt} & = & f_{{\rm in}i}
({\cal R}X_i+{\cal Y}_i)
-[\alpha -1+{\cal R}-\beta_{\rm acc}(1-f_i)+\beta_{\rm SN}]
{\cal D}_i,\label{basic6}
\end{eqnarray}
where ${\cal D}_i\equiv M_{{\rm d}i}/M_{\rm g}=f_iX_i$, and
$\beta_{\rm acc}$ and $\beta_{\rm SN}$ are, respectively,
defined by
%\begin{eqnarray}
$\beta_{\rm acc}\equiv{M_{\rm g}}/{(\tau_{\rm acc}\psi )}$
{and} $\beta_{\rm SN}\equiv{M_{\rm g}}/{(\tau_{\rm SN}\psi )}$.
%\label{betas}
%\end{eqnarray}
(See also Eqs. 10--12 in LF98.) We can regard $\beta_{\rm SN}$
as constant in time (\S 6.2 of D98). The reason is as follows. The
destruction rate
of dust by SNRs %, $[dM_{{\rm d}i}/dt]_{\rm SNR}$ 
is written as
%\begin{eqnarray}
%\left[\frac{dM_{{\rm d}i}}{dt}\right]_{\rm SNR}=
$({M_{{\rm d}i}}/{M_{\rm g}})\epsilon M_{\rm SNR}R_{\rm SN}$,
%\end{eqnarray}
where $\epsilon$ is the grain destruction efficiency, $M_{\rm SNR}$
is the total mass of interstellar gas swept up by a SNR during its
lifetime, and $R_{\rm SN}$ is the number of SNe (Type I and Type II)
per unit time. Since this should be equivalent to the last term of
right-hand side of equation (\ref{basic3}),
$\tau_{\rm SN}={M_{\rm g}}/(\epsilon M_{\rm SNR}R_{\rm SN})$.
Because the dependence of $\epsilon M_{\rm SNR}$ on the gas
density is weak (Jones, Tielens, \& Hollenbach 1996), we regard
$\epsilon M_{\rm SNR}$
as a constant. If we reasonably assume that $R_{\rm SN}$
is proportional to the star formation rate, we derive
$\tau_{\rm SN} \propto {M_{\rm g}} / {\psi}$.
This is why $\beta_{\rm SN}=M_{\rm g}/(\tau_{\rm SN}\psi)$ can
be treated as a constant.
The  Galactic value shows  $\beta_{\rm SN}\sim 5$ (LF98).
This value corresponds
to $\tau_{\rm SN}\sim 10^8$ yr, which is consistent with
Jones et al. (1994) and
Jones, Tielens, \& Hollenbach (1996).
The relation $\tau_{\rm acc}\simeq\tau_{\rm SN}/2$ (D98) leads to
$\beta_{\rm acc}\simeq 2\beta_{\rm SN}\simeq 10$.

Combining equations (\ref{basic5}) and (\ref{basic6}),
we obtain the following differential equation of ${\cal D}_i$
as a function of $X_i$:
\begin{eqnarray}
{\cal Y}_i\frac{d{\cal D}_i}{dX_i}= f_{{\rm in}i}
({\cal R}X_i+{\cal Y}_i)
-(\alpha -1+{\cal R}-\beta_{\rm acc}+\beta_{\rm SN})
{\cal D}_i-\frac{\beta_{\rm acc}{\cal D}_i^2}{X_i},
\label{difeq}
\end{eqnarray}
where we used the relation of $f_{ i}={\cal D}_i/X_i$.
%The above equation is integrated by the Runge-Kutta
%method.

\subsection{Comparison with Observations}

The result of the numerical integration of equation (\ref{difeq})
is compared with  observational data of nearby
spiral galaxies. 
We take again $i={\rm O}$ (see also \S 2.1).
Moreover, we make an assumption that total mass of dust 
including both oxygen dust and carbon dust\footnote{We should note
that the carbon dust is not traced by oxygen.} is 
proportional to that of oxygen in the dust phase.
In other words, 
\begin{eqnarray}
 {\cal D}\equiv\sum_i{\cal D}_i=C
 {\cal D}_{\rm O},
\end{eqnarray}
where $C$ is assumed to be constant for all spiral galaxies.
We adopt the Galactic value, $C\simeq 2.2$ according to Table 2.2 of
Whittet (1992).

For further quantification, we need to fix the values of
${\cal R}$ and ${\cal Y}_i$ for the traced element ($i={\rm O}$).
%LF98 provided the values of the two quantities for the oxygen.
We choose oxygen as the tracer, because
%The reason why LF98 chose oxygen as the tracer is as follows:
(i) Most of it is produced in type II SNe which are
also responsible for the shock destruction of dust grains; (ii)
oxygen is the main constituent of dust grains (see also LF98).
The point
(i) means that the instantaneous recycling approximation may
be a reasonable approximation for the investigation of oxygen
abundances, since the generation of oxygen is a
massive-star-weighed phenomenon. In other words, results are
insensitive to the value of $m_{\rm l}$ defined in equations
(\ref{eqr}) and (\ref{eqy}).
According to LF98, we put
$m_{\rm l}=1M_\odot$ and $m_{\rm u}=120M_\odot$.
We use a power-law form of the IMF: $\phi(m)\propto m^{-x}$.
Two kinds of the IMFs are investigated: $x=2.35$ (Salpeter 1955) and
$x=2.7$ (Scalo 1986). If we assume that the lower cutoff of the
stellar mass is equal to $m_{\rm l}(=1M_\odot )$,
$({\cal R},~{\cal Y}_{\rm O})=
(0.79,~1.8\times 10^{-2})$ and
$(0.68,~9.6\times 10^{-3})$ for the IMFs of Salpeter and Scalo,
respectively (LF98). Adopting the lower cutoff of $0.1M_\odot$
instead, the values become $({\cal R},~{\cal Y}_{\rm O})
=(0.32,~7.2\times 10^{-3})$ and $(0.13,~1.8\times 10^{-3})$
for the Salpeter and Scalo IMFs, respectively.
We use the former values for ${\cal R}$ and ${\cal Y}_{\rm O}$.
We treat the latter values only in \S 2.2.3.

\subsubsection{Dependence on $\beta_{\rm acc}$}

First, we examine the dependence of $X_{\rm O}$--${\cal D}$
relation on the parameter $\beta_{\rm acc} $,
the efficiency of accretion
of heavy element onto the preexisting dust grains.
The result is shown in Figure 1.

The data points are from Issa,  MacLaren, \& Wolfendale (1990;
see also the references therein).
The dust-to-gas ratio is calculated from the interstellar
extinction and column densities of H {\sc i} and H$_2$.
Each value is  evaluated at a galactocentric distance equivalent
to that of the Sun (the galactocentric distance of each
galaxy is properly scaled according to its size). 

We show the cases of
$\beta_{\rm acc}=5,~10,$ and 20 (the solid, dotted, and dashed
lines, respectively),  with the relation
$\beta_{\rm acc}=2\beta_{\rm SN}$ fixed.
The increase of
$\beta_{\rm acc}$ means that
accretion of heavy elements onto preexisting dust becomes efficient.
Thus, for a fixed value of the metallicity, dust-to-gas
ratio increases as $\beta_{\rm acc}$ increases.
We also show the case of $\beta_{\rm acc}=0$ and $\beta_{\rm SN}=10$
(long-dashed line), which is far from reproducing the data. This means
that
%we cannot reproduce the observation unless
we should take into account the accretion onto the preexisting dust
as Dwek \& Scalo (1980) asserted.
Thus, the models in LF98, which do not include the term
of the accretion, cannot be applied to spiral galaxies.
The equations in LF98
can be applied to dwarf galaxies, where the accretion process may be
inefficient because of their low metallicity. The accretion process is
properly considered in D98.

\subsubsection{Dependence on $f_{{\rm in}i}$}

 In Figure 2, the solid, dotted, and dashed lines show the
${\cal D}$--$X_{\rm O}$ relations for $f_{\rm inO}=0.4$, 0.1, and
0.01, respectively (in the case of
$\beta_{\rm acc}=2\beta_{\rm SN}=10$ and the Salpeter IMF).
In LF98, $f_{\rm inO}\simeq 0.4$ is
suggested to be an optimistic upper limit. Figure 2 shows that
the larger efficiency of metal production 
from heavy elements leads to the larger dust-to-gas ratio.

In the limit of $X_{\rm O}\to 0$, the solution of equation
(\ref{difeq}) reduces to ${\cal D}\simeq Cf_{\rm inO}X_{\rm O}$
(see also LF98), so that ${\cal D}$ scales linearly with
$f_{\rm inO}$. Thus, the precise value of $f_{\rm inO}$ can be
constrained by low-metal galaxies. Indeed,
the difference among the three cases is significant in the
low-$X_{\rm O}$ region in Figure 2. It seems unlikely that
$f_{\rm inO}\lnyoro 0.01$, since the data of the low-metal spirals
cannot be reproduced by such small $f_{\rm inO}$.

\subsubsection{Dependence on IMF}

The difference of mass function is represented by different
${\cal R}$ and ${\cal Y}_{\rm O}$ (\S 2.1). We examine
how the difference affects the
result. We present the result in Figure 4. The Salpeter
and Scalo IMFs are represented by the solid and dotted lines,
respectively (lower ones). As commented above, if we use the lower
mass cutoff for the stellar
mass, ${\cal R}$ and ${\cal Y}_{\rm O}$ become
smaller. We also show the case of lower mass cutoff ($0.1M_\odot$;
the upper line for each IMF).
The differences among the results are too small (comparable to the
typical observational errors) to determine which line is preferable.

\section{SUMMARY AND CONCLUSIONS}

Based on the models proposed by LF98 and D98,
we have examined the dust content in nearby spiral galaxies.
We simplified the basic equations in D98 and applied the
instantaneous recycling approximation to them. We treated
one-zone model equations
which describe the changing rate of dust-to-gas
ratio including the terms of dust condensation from heavy
elements ejected by stars, destruction by SN shocks,
destruction in star-forming regions, and
accretion of elements onto preexisting dust grains (\S 2.1).
Assuming that the  total dust mass
is proportional to
the mass of oxygen in the dust phase, we compared
the results\footnote{We should keep in mind the
uncertainty concerning the observational data (e.g., the uncertainty in
CO-to-H$_2$ mass
conversion factor; Arimoto, Sofue, \& Tsujimoto 1996).
} with the data of nearby spiral
galaxies (for the detailed comparison with the Galactic values,
see D98, in which the data are excellently reproduced).
Finally, we summarize the conclusions reached in this paper:

(i)
Unless we do not take into account the accretion process of heavy
elements onto the preexisting dust particles, 
 we cannot explain the
observed relations between dust-to-gas ratios and metallicities
of nearby spiral galaxies. The accretion
process is important for high-metallicity systems, since the
collision rate between heavy-element atoms and dust
grains is large. The data of
dwarf galaxies, which are generally low-metallicity
systems, may be explained even if the accretion process is
neglected (LF98).
%Indeed, LF98 explained the relation between dust-to-gas ratio and
%metallicity without considering the accretion process.
The recent model by D98 properly treated the accretion.

(ii) The efficiency of dust production  from heavy element
(denoted by $f_{{\rm in}i}$)
can be constrained by the galaxies with low metallicity.
It is unlikely that $f_{{\rm in}i}\lnyoro 0.01$.

(iii) As for IMFs, both the Salpeter and Scalo types are
consistent with the observed relation between dust-to-gas ratio and
metallicity.

\acknowledgements
We are grateful to S. Mineshige for continuous
encouragement.
We wish to thank E. Dwek for many comments which improved
discussions in this paper and our knowledge of recent
studies on dust. We acknowledge the Research
Fellowship of the Japan Society for the
Promotion of Science for Young Scientists.

\newpage

\centerline{\bf FIGURE CAPTIONS}

\noindent
FIG. 1---
The relations between dust-to-gas mass ratio (${\cal D}$) and
oxygen abundance (oxygen mass fraction $X_{\rm O}$)
for various  $\beta_{\rm acc}$ and $\beta_{\rm SN}$
(dust accretion efficiency onto the preexisting dust grains and
dust destruction efficiency in the SN shocks, respectively),
with the relation $\beta_{\rm acc}=2\beta_{\rm SN}$ kept.
The data points for nearby spiral galaxies are from Issa, 
 MacLaren, \& Wolfendale (1990). Here,
$f_{\rm inO}$ is fixed to 0.1, and the Salpeter IMF is adopted.
The solid, dotted, and dashed lines represent the cases
of $\beta_{\rm acc}=2\beta_{\rm SN}=5$, 10, and 20, respectively.
The long-dashed line shows the case of no accretion process onto
the preexisting dust grains ($\beta_{\rm acc}=0$ and
$\beta_{\rm SN}=5$).

\medskip

\noindent
FIG. 2---
The same as Fig.~1 but for various $f_{\rm inO}$ (condensation
efficiency of oxygen into dust grains). The other parameters are
set to
$\beta_{\rm acc}=2\beta_{\rm SN}=10$.
The Salpeter IMF is adopted. The solid, dotted, and dashed lines
represent the cases
of $f_{\rm inO}=0.5,~0.1$, and 0.01, respectively.

\medskip

\noindent
FIG. 3---
 The same as Fig.~1 but for the different IMFs (Salpeter and
Scalo; solid and dashed lines, respectively).
Two lower cutoffs of
stellar mass are investigated for each IMF: the cutoff of
$1M_\odot$ and $0.1M_\odot$ (the lower and upper lines,
respectively). 
The parameters are fixed to $\beta_{\rm acc}=2\beta_{\rm SN}=10$
and $f_{\rm inO}=0.1$.

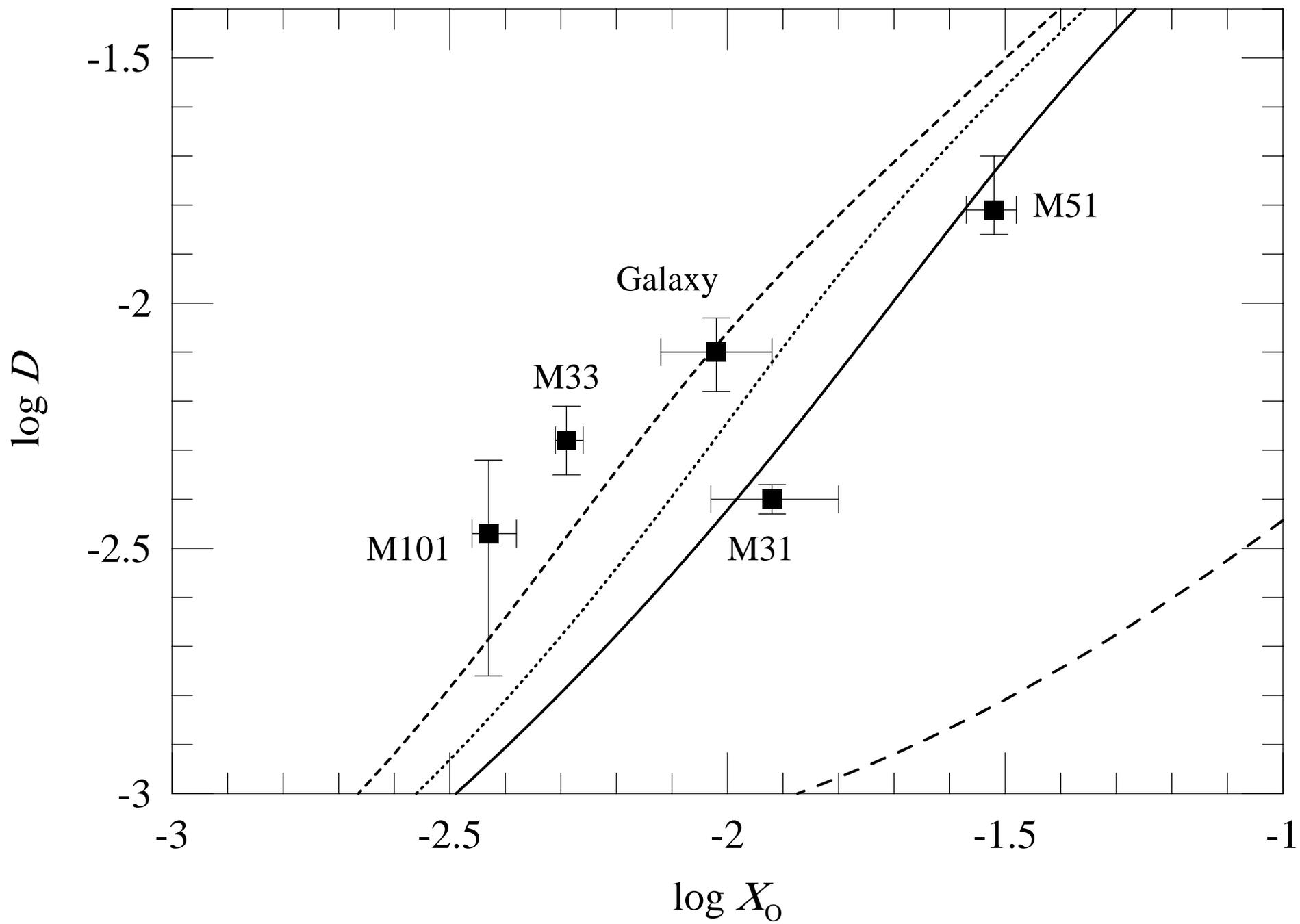
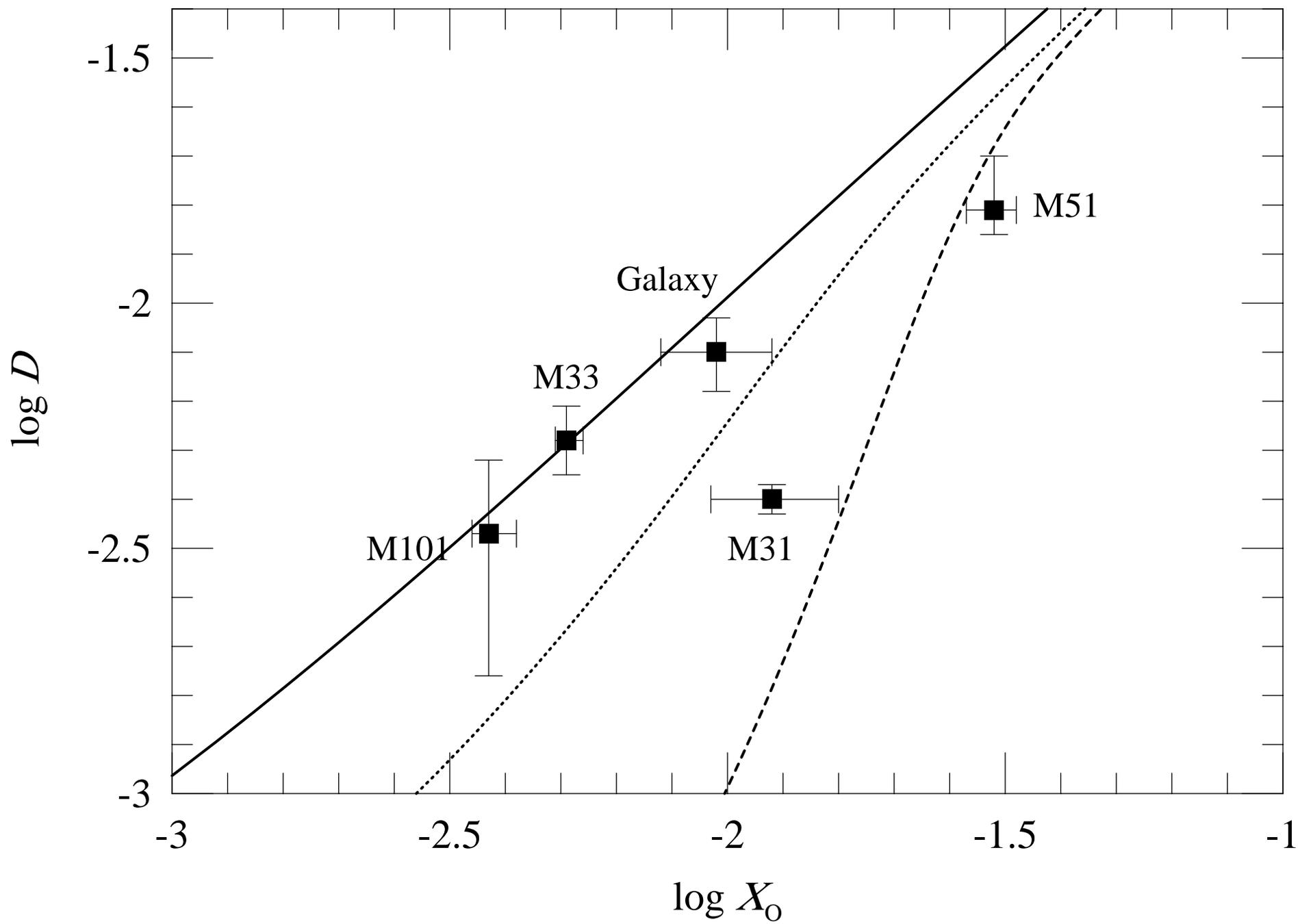
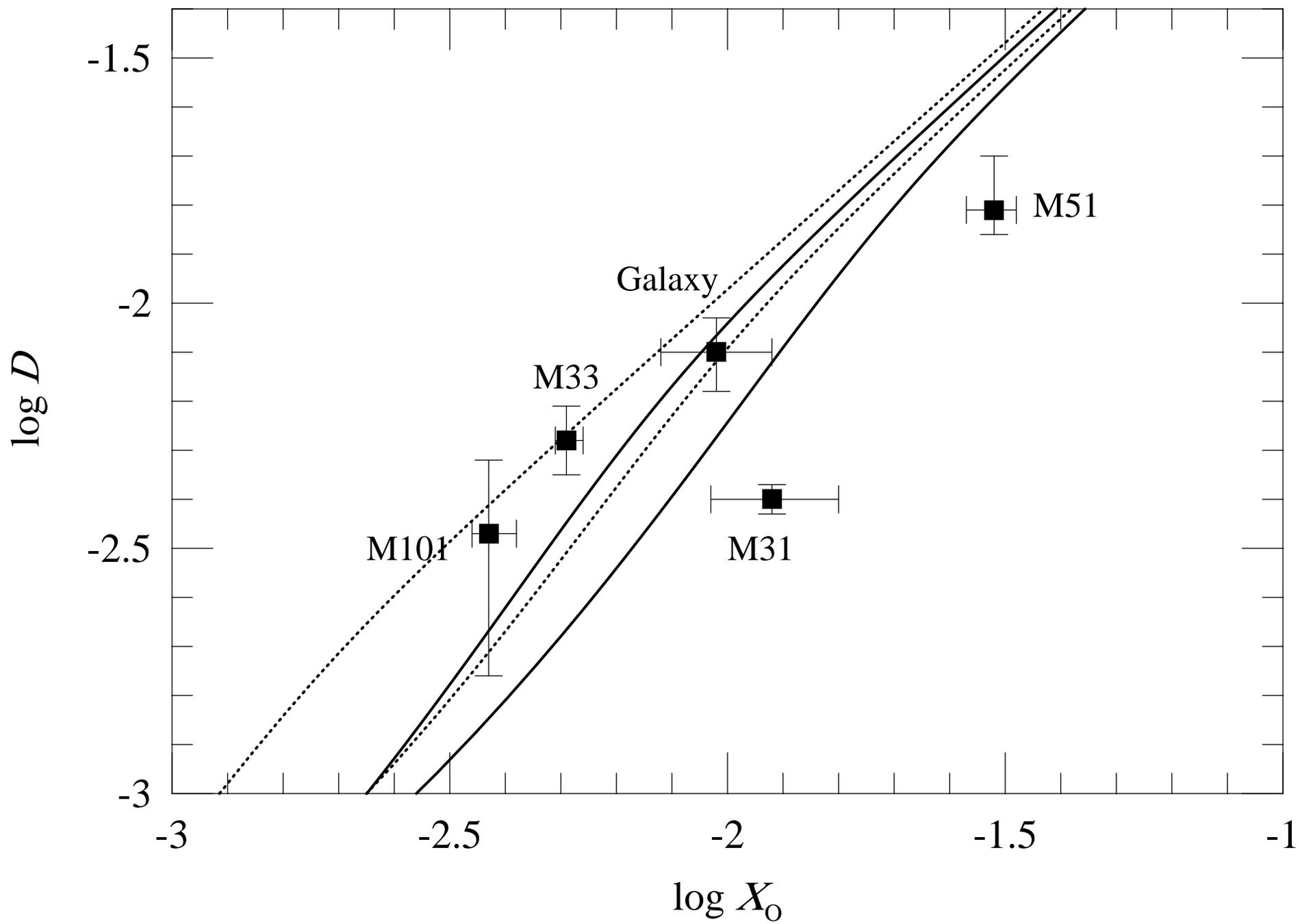
\end{document}